\documentclass[iop]{emulateapj}
\begin{document}

\title{Resolved Magnetic Field Mapping of a Molecular Cloud using GPIPS}
\author{Robert C. Marchwinski, Michael D. Pavel, \& Dan P. Clemens}
\affil{Institute for Astrophysical Research}
\affil{Boston University, 725 Commonwealth Ave, Boston, MA 02215}
\email{robmarch@bu.edu; pavelmi@bu.edu; clemens@bu.edu}

\slugcomment{Published in The Astrophysical Journal}

\shorttitle{GPIPS Probe of GRSMC 45.60+0.30}
\shortauthors{Marchwinski, R. C., Pavel, M. D., Clemens, D. P.}

\begin{abstract}
We present the first resolved map of plane-of-sky magnetic field strength for a quiescent
molecular cloud. GRSMC 45.60+0.30 subtends 40~$\times$~10 pc at a distance of 1.88~kpc, masses 16,000~$M_{\sun}$, and exhibits no star formation. Near-infrared background starlight polarizations were obtained for the Galactic Plane Infrared Polarization Survey using the 1.8m Perkins telescope and the Mimir instrument. The cloud area of 0.78 square degrees contains 2,684 significant starlight polarizations for 2MASS-matched stars brighter than 12.5~mag in $H$-band. Polarizations are generally aligned with the cloud's major axis, showing an average P.A. dispersion of $15^{\circ}\pm2^{\circ}$ and polarization of $1.8\pm0.6\%$. The polarizations were combined with Galactic Ring Survey $^{13}$CO spectroscopy and the Chandrasekhar-Fermi method to estimate plane-of-sky magnetic field strengths, with an angular resolution of 100 arcsec. The average plane-of-sky magnetic field strength across the cloud is $5.40\pm0.04$~$\mu$G. The magnetic field strength map exhibits seven enhancements, or `magnetic cores.' These cores show an average magnetic field strength of $8.3\pm0.9$~$\mu$G, radius of $1.2\pm0.2$~pc, intercore spacing of $5.7\pm0.9$~pc, and exclusively subcritical mass-to-flux ratios, implying their magnetic fields continue to suppress star formation. The magnetic field strength shows a power law dependence on gas volume density, with slope $0.75\pm0.02$ for $n_{H_2}$~$\geq10$~cm$^{-3}$. This power law index is identical to those in studies at higher densities, but disagrees with predictions for the densities probed here.

\end{abstract}
\keywords{ISM: magnetic fields - polarization - molecular clouds - Individual Object: GRSMC 45.60+0.30}

\section{Introduction}

What roles do magnetic fields play in molecular clouds? Within clouds, important forces include gravity, gas pressure, cosmic ray pressure, and magnetic fields; however, they are not independent of each other. For example, \citet{HC05} point out that cosmic rays and gas pressure are coupled through the magnetic field. Therefore, the magnetic field may be a major factor in cloud dynamics across many scales and may be a key agent in regulating the rate of star formation \citep{FIE65}. Yet, the magnetic field is difficult to sense and harder still to map in much detail.

In the hot ionized interstellar medium, magnetic fields are probed in diffuse regions through Faraday rotation of background pulsars \citep{SM68,MAN74,HAN06} and extragalactic sources \citep{COO62,PS11}, as well as through synchrotron emission \citep{WES62,WIE62,YUS84,YUS89}. In cold, denser regions, the magnetic fields are probed through Zeeman splitting of spectral lines \citep{VER68,CRUT3}, polarized thermal emission from aligned dust \citep{HIL88,DOT10}, and polarization of background starlight by aligned dust \citep{MF70,HEI00}. Zeeman splitting and Faraday rotation measure the line-of-sight magnetic field strength; the other methods trace the orientation of the magnetic field on the plane of the sky. Chandrasekhar \& Fermi (1953a; hereafter CF) derived an estimate of the plane-of-sky magnetic field strength using the dispersion of the starlight polarization position angles (P.A.s), the local gas motions, and the local gas density. 

Polarization of background starlight, first observed by \citet{HILT} and \citet{HALL}, has been attributed to elongated dust grains aligned by local magnetic fields. Early alignment mechanisms were proposed by \citet{DG51}, \citet{GOLD}, and \citet{PS71}. \citet{L03} provides a comprehensive history of grain alignment.  Currently, the favored model for dust alignment is the radiative torque mechanism \citep{DG76,L03,L07}. With radiative torques, spinning anisotropic dust grains preferentially align their long axes perpendicular to the local magnetic field. Unpolarized background starlight sees a higher extinction cross-section perpendicular to the local magnetic field direction and the light picks up a small linear polarization parallel to the projected magnetic field direction. With this mechanism operating, magnetic fields can be probed within molecular clouds using optical or near-infrared polarimetric observations of background stars.

The Galactic Plane Infrared Polarization Survey \citep[GPIPS;][]{DAN11}, from which these data are drawn, is a near-infrared, linear polarization survey that measures the polarization of background starlight in the inner Galactic mid-plane ($18\degr < \ell < 56\degr$, $-1\degr < b < 1\degr$), to and beyond the nearest spiral arm. This survey obtains $H$-band (1.6~$\mu$m) linear polarizations for apparent magnitudes from 7th to beyond 14th. This same Galactic region has already been surveyed by $IRAS$ \citep{NEU84}, NVSS \citep{NVS98}, $MSX$ \citep{PRI01}, GLIMPSE \citep{GLIMP}, the $^{13}$CO Galactic Ring Survey \citep[GRS;][]{JAC06}, 2MASS \citep{2MASS}, MIPSGAL \citep{CAR09}, $WISE$ \citep{WRI10}, and BGPS \citep{BCS11}, and so offers excellent data sets for correlative analysis.

The newly available GPIPS background starlight polarimetry, with its roughly one~arcmin stellar sampling, is ideal for exploring the nature of magnetic fields within molecular clouds. A key first step is to measure and characterize the magnetic field within an average molecular cloud that is not engaged in active star formation. With GPIPS polarimetric data, combined with velocity information from GRS, the CF method can be used to create a resolved map of a cloud's embedded plane-of-sky magnetic field strength. Such maps can be analyzed to examine both structural properties and key diagnostic relationships underlying the physical conditions in the cloud.

\begin{deluxetable*}{cccccc}
\tabletypesize{\footnotesize}
\centering
\tablewidth{0pt}
\tablecolumns{6} 
\tablecaption{Cloud Properties}
\tablehead{\colhead{Cloud Desig.} &
\colhead{Identifier} & \colhead{Distance\,\tablenotemark{a}} & \colhead{$V_{LSR}$ Range\,\tablenotemark{b}} &
\colhead{Extent} &\colhead{Mass\,\tablenotemark{b}} \\
\colhead{} &\colhead{} &\colhead{[kpc]} &\colhead{[km s$^{-1}$]} &\colhead{[pc]} &\colhead{[$M_{\sun}$]} }
\startdata 
GRSMC 45.60+0.30 & Cloud 1 & 1.88 & 22.0 - 30.0 & $45 \times 22$ & $17,000$  \\
GRSMC 45.46+0.05 & Cloud 2 & 7.45 & 50.0 - 66.0 & $75 \times 53$ & $49,000$  \\
\enddata
\label{tab:cloud}
\tablenotetext{a}{\citet{DUV09}}
\tablenotetext{b}{\citet{SIM01}}			 
\end{deluxetable*}

\begin{figure}
\plotone{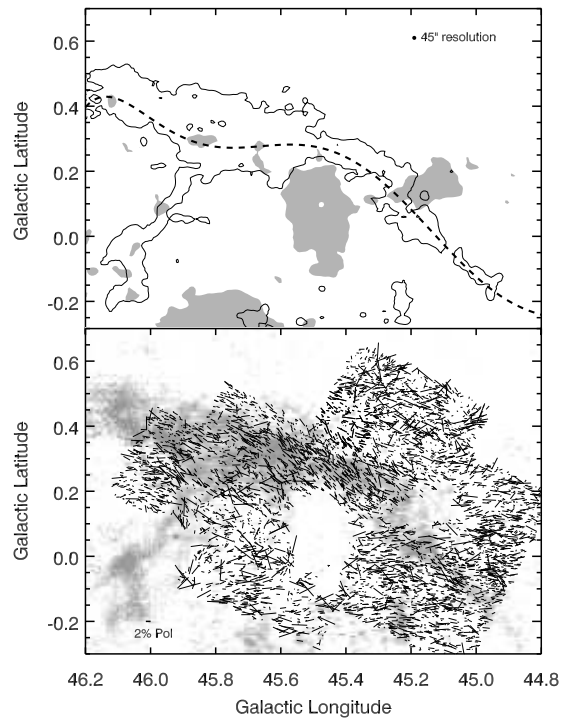}
\caption{Map showing extents and locations of Cloud~1 (quiescent) and Cloud~2 (active). Cloud~1 is delineated by a single $^{13}$CO contour representing peak antenna temperature at the 1.0 K level (black contour). Cloud~2 is shown as the filled, gray region, representing $^{13}$CO peak antenna temperature at and above the 6 K level. The bold, dashed curve traces the `spine' of Cloud 1, as discussed in Section 2.3. (\emph{Bottom}) Gray scale image of Cloud~1 $^{13}$CO integrated line intensity (between $V_{LSR}$ 20 and 32 km s$^{-1}$). Overlaid are the high-quality, 2MASS-matched GPIPS polarization vectors assigned to Cloud~1 (vectors overlapping Cloud~2 have been removed). Vector lengths are proportional to the polarization percentages, with a 2\% reference vector shown in the lower left corner. Vector orientations show the polarization position angles (directions of the projected maximum of the electric field vector). The GRS data have 45 arcsec resolution, represented by the circle in the upper right.}
\label{fig:Cooverlay_paper}
\end{figure}

A region near $\ell=45\degr.5,b=+0\degr.2$ was chosen for this quiescent cloud study. In this direction, GRS data reveal the presence of two coincident molecular clouds \citep{SIM01} separated in velocity, and therefore space. These two clouds are the quiescent cloud GRSMC 45.60+0.3, hereafter `Cloud~1,' and the actively star-forming cloud GRSMC 45.46+0.05, hereafter `Cloud~2' (see Figure \ref{fig:Cooverlay_paper} top). \citet{SIM01} and \citet{DUV09} established general properties for both clouds, as summarized in Table \ref{tab:cloud}. \citet{DUV09} used {H\kern0.1em{\sc i}} spectral line self-absorption to resolve the near-far kinematic distance ambiguity to place Cloud~1 at a heliocentric distance of 1.88~kpc (near) and Cloud~2 at 7.45~kpc (far).

%% Fig 1

This work explores the magnetic field structure toward this region by combining GPIPS starlight polarimetry, GRS $^{13}$CO spectra, and 2MASS photometry. By correlating extinction maps with integrated $^{13}$CO images over each cloud's velocity range, the observed stellar color excess, and therefore background starlight polarimetry, was found to be associated primarily with the quiescent Cloud~1. The CF method was used to estimate the strength of the magnetic field, in the plane of the sky, across the full extent of Cloud~1, with 100~arcsec angular resolution. These results were used to create, for the first time, a resolved map of the magnetic field strength across a molecular cloud and to analyze the contents of this map. In Section~2, the cloud selection and its properties are described. Section~3 discusses the GPIPS observations and Section~4 describes the data analysis. The results are presented in Section~5 and Section~6 contains a discussion of these results. In Section 7, the work is summarized.  

\section{Extinction Mapping and $^{13}$CO Data Analysis}

The sky region chosen contains both a quiescent and an active star forming cloud. In the following, the polarization properties traced by GPIPS are shown to be associated with Cloud~1, the more nearby and quiescent cloud. Cloud~1's structural properties are then explored.

\subsection{Extinction Mapping}

To quantify which cloud is contributing to the extinction, and therefore the NIR polarization of each observed star, the 2MASS color excess $E(H-K)$, a proxy for extinction, was compared to the $^{13}$CO intensity across each of the two clouds. Visual extinctions were estimated as:
\begin{equation} \label{eq:one}
A_{\mbox{v}} \approx r^{\prime} E(H-K),
\end{equation}
where r$^{\prime}$ is approximately 12-13 for molecular clouds \citep{WH03}. For every 2MASS-matched star within our polarimetric sample (see Section 3), $E(H-K)$ was calculated using:
\begin{equation} \label{eq:two}
E(H-K) = (H-K)_{2MASS}-<(H-K)_{intrinsic}>
\end{equation}
where $<(H-K)_{intrinsic}>$ is $0.13\pm0.01$~mag \citep{LAD94}. The GRS spectral-spatial 3-D data have 45~arcsec angular resolution and are presented as spectra for a Nyquist-sampled, 22~arcsec spatial grid. Using the same spatial grid, the 2MASS-based extinctions for all stars having measured polarizations (see Section~3) and within 150 arcsec of each grid center were averaged, using gaussian weighting by angular offset from the grid centers, as well as by color uncertainties, to yield mean stellar extinctions $<$A$_{\mbox{v}}$$>$. 

\begin{figure}[b]
\centering
\includegraphics[angle=90,scale=0.39]{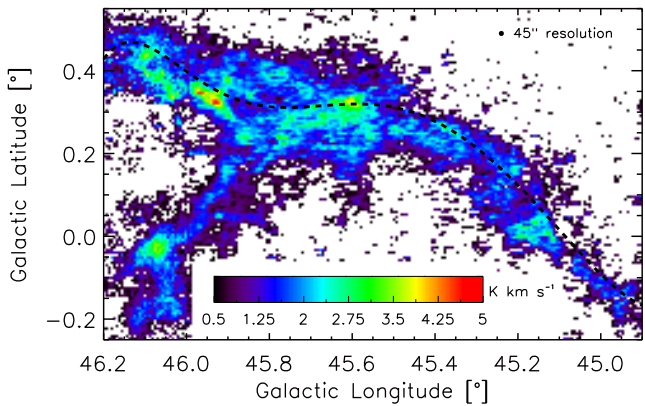}
\caption{$^{13}$CO integrated intensity over the velocity limits appropriate to Cloud 1 (see Table \ref{tab:cloud}). The color bar represents integrated intensity values from 0.5 to 5 K km s$^{-1}$. The GRS data have 45~arcsec resolution, represented by the filled circle in the upper right.}
\label{fig:coint}
\end{figure}

\begin{figure}[b]
\includegraphics[angle=90,scale=0.35]{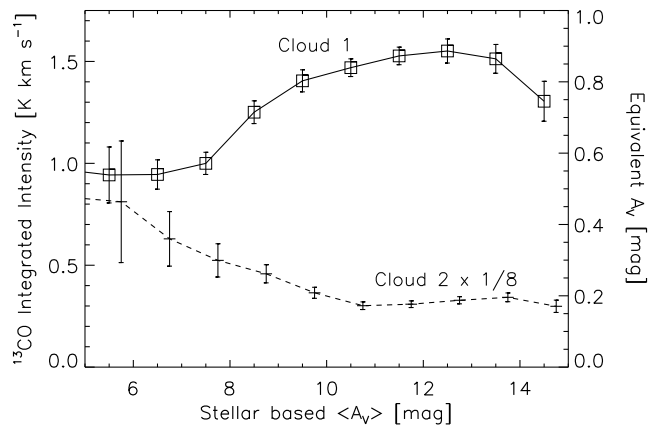}
\caption{Averaged $^{13}$CO integrated intensity found in bins of mean stellar extinction $<$A$_V$$>$ for Cloud 1 (solid line) and Cloud 2 (dashed line). For Cloud 2, the $^{13}$CO intensity has been reduced by a factor of eight. Extinction correlates with $^{13}$CO intensity above $<$A$_V$$>$$ \sim 6$ mag for Cloud 1, but not for Cloud 2.}
\label{fig:avco}
\end{figure}

To test the degree to which the stellar extinctions are associated with each cloud, the mean extinctions were compared to the GRS integrated $^{13}$CO intensities arising from each cloud at each spatial grid point. The $^{13}$CO emission for each cloud was integrated across the velocity ranges listed in Table \ref{tab:cloud}. For example, the Cloud 1 integrated intensity map is presented as Figure \ref{fig:coint}. Direct comparisons of the mean stellar extinctions and $^{13}$CO integrated intensities for each grid center were dominated by noise, so these data were binned by $<$A$_{\mbox{v}}$$>$ and are shown as Figure~\ref{fig:avco}. In this Fig., the solid line reveals a relationship between mean stellar extinction and $^{13}$CO intensity (and equivalent $A_V$) for Cloud~1, while the dashed line shows no such relationship for Cloud~2. Hence, the dust extinction, and therefore the NIR polarizations, are primarily associated with Cloud~1. Further, to avoid any possible polarization contamination arising from Cloud~2, all stars falling within regions showing Cloud~2 $^{13}$CO integrated intensity values larger than 6~K~km~s$^{-1}$ were removed from further consideration (see Fig.~1 bottom, and discussion below). This conservative cut excludes contamination by Cloud~2 and allows the magnetic field of Cloud~1 to be reliably probed.

%% Fig 2

%% Fig 3

\subsection{Quiescent Nature of Cloud 1}

To establish the quiescent nature of Cloud~1, the GLIMPSE, $IRAS$, MIPSGAL, $MSX$, $WISE$, and BGPS data for this region were examined. No BOLOCAM point sources appeared within the cloud, and there was no increased density of GLIMPSE targets, whose presence might signify recently formed stars or clusters. No extended 4.5~$\mu$m emission was seen, also suggesting there is no active star formation \citep{CHAM09}. Furthermore, no $IRAS$, MIPSGAL, $MSX$, or $WISE$ emission sources were seen within Cloud~1 at wavelengths that might signify active star formation. 

Recently, \citet{LAD10} showed that the presence of star formation seems to correlate with cloud-based extinctions of A$_{\mbox{v}}~>~7$ mag. To test for the presence of high
extinction zones in Cloud 1, gaussians were fit to the GRS $^{13}$CO spectra across the 
cloud's angular and velocity extents (Table \ref{tab:cloud}), returning the peak antenna temperature and FWHM velocity width for each GRS spatial pixel.
Using:
\begin{equation} \label{eq:three}
N_{H_{2}} = (4.92 \times 10^{20}) \: T \: \sigma_{v} \: \: \: \: \: [cm^{-2}]
\end{equation}
from \citet{SIM01}, where T is the peak antenna temperature in K, and $\sigma_{v}$ is the line width in km s$^{-1}$, the column density across the cloud was obtained.  These column density values were transformed to $<$A$_{\mbox{v}}$$>$ using:
\begin{equation} \label{eq:duh}
<A_{\mbox{v}}> \: =  1.06 \times 10^{-21} \: N_{H_2} \: \: \: \: \: [mag]
\end{equation}
from \citet{PIN08}, yielding estimated cloud extinctions. The mean extinction across the cloud is about 0.4 mag at V and nowhere reaches more than 2.6 mag,  supporting the conclusion that Cloud~1 is quiescent. Differencing the mean ($H - K$) stellar colors
for a $20 \times 20$\arcmin\, region centered at ($\ell$, $b$)~$\sim$~(45\degr.3, $+$0\degr.5) off Cloud 1 with the
mean ($H - K$) colors of stars on Cloud 1 also returns a mean cloud-induced $A_V$ of about 2.4 mag for the polarization stars.

\subsection{Cloud 1 Gas-traced Properties}

\begin{figure}
\centering
\includegraphics[scale=0.39,angle=90]{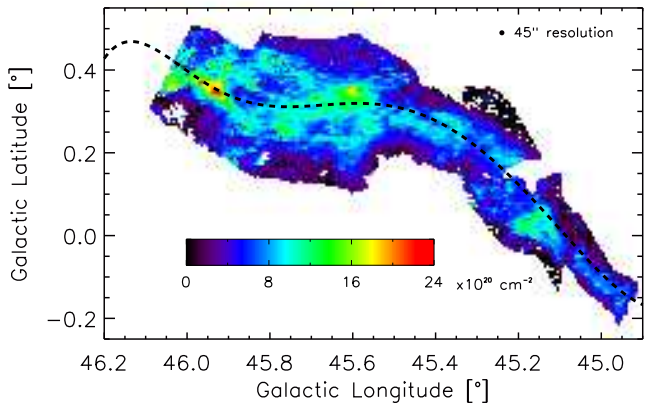}
\caption{H$_2$ column density, calculated from GRS data using Eq~(\ref{eq:three}), across Cloud 1. The color bar represents values from 0 to $24\times10^{20}$ cm$^{-2}$. The GRS data have 45 arcsec resolution, represented by the filled circle in the upper right.}
\label{fig:colden}
\end{figure}

To use the CF method, the gas volume density all across Cloud~1 must be estimated. Observationally, the mean gas volume density is obtained from measuring the gas column density and the cloud thickness, or depth, along the line of sight. The column density of molecular hydrogen, N$_{H_{2}}$, was estimated using Eq (\ref{eq:three}), which was used to create a column density map across the cloud, shown as Figure \ref{fig:colden}. (The truncated cloud boundaries in the top left and lower left are a result of GPIPS having not yet observed those fields for $H$-band polarization.)

%% Fig 4

The cloud thickness was estimated by assuming cylindrical symmetry, in that the plane-of-sky projected cloud minor axis diameter serves as a proxy for the full line-of-sight thickness `$\Delta$z.' To begin characterizing the spatial variations of cloud thickness, the central axis or `spine' of the cloud was identified. The $^{13}$CO data cube for Cloud~1 was integrated over the cloud's velocity range (Table \ref{tab:cloud}) and was spatially boxcar-smoothed (at 10 pixels). For each (constant $\ell$) column in the resulting integrated intensity image, the spatial ($b$) pixel with the highest integrated intensity was identified. These pixels were presumed to trace the cloud's central axis, were simply connected, and were found to lie near the center of the cloud's plane-of-sky projection. However, because these points were not found to be linearly distributed on the sky, a polynomial fit of these high intensity points (in $\ell$ and $b$) was used to model the location of the `spine' of the cloud. An F-test revealed the best fit was a seventh-order polynomial, shown as the dashed line in the top panel of Fig. \ref{fig:Cooverlay_paper}, in Fig.~\ref{fig:colden}, and in following figures.

With the `spine' of the cloud identified, the spatial variation of the projected minor axis of the cloud along the curved spine needed to be determined. At each point along the spine $s(\ell, b)$, the plane-of-sky minor axis diameter (D) of the cloud along a line of constant Galactic longitude was measured. If the spine had been oriented perfectly along a line of constant Galactic latitude, this diameter would be appropriate. However, the spine is generally not perfectly aligned with the Galactic coordinate system and so measuring the diameter only along $b$ will tend to overestimate the true cloud minor-axis diameter.

To correct for this, the relative orientation between the spine and the Galactic coordinate system was parametrized along the cloud length. At each point along the spine, the angle $\theta(s)$ between a normal to the spine and the direction to Galactic North ($+b$ direction) was calculated. An estimate of the true cloud diameter was then:
\begin{equation}
 D'(s) = D(s)*cos(\theta(s)).
 \end{equation}

\begin{figure}
\centering
\includegraphics[scale=0.39,angle=90]{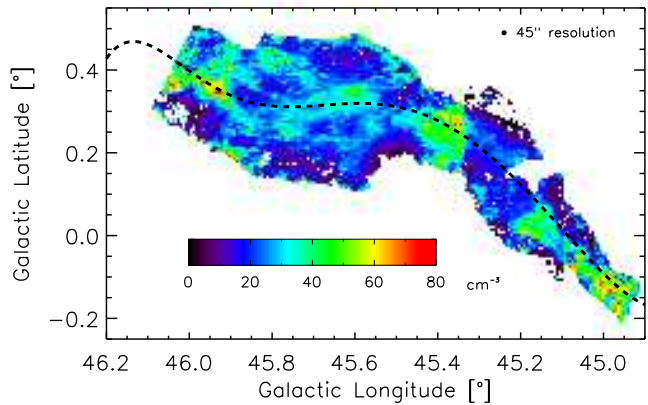}
\caption{Mean H$_2$ gas volume number density, calculated from GRS data using Eq~(\ref{eq:five}), across Cloud 1. The color bar represents number density values from 0 to 80~cm$^{-3}$. The GRS data have 45~arcsec resolution, represented by the filled circle in the upper right.}
\label{fig:numden}
\end{figure}

With knowledge of the spine's location and the variation of the corrected minor-axis diameter of the cloud (D$'$(s)), the line-of-sight thickness through the cloud could be calculated. For a given direction through the cloud, the shortest distance to any point on the spine was calculated (d) and the estimated cloud diameter D$'$(s) was identified for that spine location. The line-of-sight thickness becomes the length of a chord, at impact distance d, from the center of a circle with diameter D$'$(s):
\begin{equation} \label{eq:four}
\Delta z(\ell, b) =  2 \: \Big{[} \: \Big{(} \frac{D' \: s(\ell, b) }{2} \Big{)}^2 - d(\ell, b)^2 \: \Big{]} \: ^{0.5}
\: \: \: \: [cm]\: .
\end{equation}
Along each direction through the cloud, the H$_2$ column density N(H$_2$) was previously determined from the $^{13}$CO integrated spectra (see Fig. \ref{fig:colden}). The volume number density was then:
\begin{equation} \label{eq:five}
n(\ell, b)_{H_2} = \frac{N_{H_{2}}(\ell, b)}{\Delta z(\ell, b)} = (4.92 \times 10^{20}) \: \frac{T(\ell, b) \: \sigma_{v}(\ell, b)}{\Delta z(\ell, b)}    \: \: \: \: \: [cm^{-3}],
\end{equation}
where $\Delta$z$(\ell, b)$ is the estimated cloud thickness, in cm. Equation (\ref{eq:five}) was used to create the mean volume number density map for the cloud shown in Figure \ref{fig:numden}. Typical number densities of H$_2$ for Cloud 1 range from 10 to 80 cm$^{-3}$. These low values indicate that significant porosity for this translucent cloud must be present to yield the $^{13}$CO emission seen for these low mean volume densities.

%% Fig 5

The average H$_2$ number density across Cloud 1 is 26 cm$^{-3}$, in close agreement with \citet{SIM01}. The estimated total mass of H$_2+$He is 15,700 $M_{\sun}$, also in agreement with the value in \citet{SIM01}. 

\section{NIR Polarimetric Data}

The polarization data used for this study were obtained with the Mimir instrument \citep{DAN07} on the Perkins 1.8m telescope between 2006 June and 2007 July as part of GPIPS. Mimir conducted near-infrared $H$-band (1.6~$\mu$m) imaging linear polarimetry over $10 \times 10$ arcmin fields with a plate scale of 0.58 arcsec using a $1024^2$ Aladdin III InSb array, operated at 33.5~K. Polarimetric analysis was performed with a cold, stepping, $H$-band, compound half-wave plate (HWP) plus a fixed, cold wire-grid analyzer.

To obtain polarimetry, exposures of 2.5s duration were obtained for 16 unique HWP position angles at six sky-dither positions, yielding a total of 96 images per field. Polarization flat fields for each HWP position were taken using a lights-on/lights-off method against an illuminated flat field screen inside the closed telescope dome. Sky observations were performed on a fixed 9 $\times$ 9~arcmin grid across the GPIPS observational area. Thirty-four GPIPS fields overlapping Cloud~1 were selected from the GPIPS data set for this study.

GPIPS data were reduced and analyzed using custom IDL software, as described in \citet{DAN11}, producing stellar polarimetry, including upper limits, for 25,966 stars over the 0.78 square degree region. This polarimetric data set goes as faint as $m_H =$12.5. Of these, 3,131 stars show significant polarization detections (P/$\sigma$$_{P}\geq3$), 2,872 of these have existing 2MASS photometry, and 2,684 remained after culling stars that overlapped Cloud~2 regions. 

\subsection{Polarization Properties}

Fig. \ref{fig:Cooverlay_paper} (bottom) shows the 2,684 significant, 2MASS-matched stellar polarization vectors required to have P/$\sigma$$_{P}\geq3$ and $\sigma$$_{P}\leq5\%$, overlaid on the $^{13}$CO integrated image for Cloud 1.
The most notable polarization features in Fig. \ref{fig:Cooverlay_paper} are the high degrees of polarization and low P.A. dispersions seen along the `spine' of Cloud 1's $^{13}$CO emission. 

\begin{figure}
\centering
\includegraphics[scale=0.35,angle=90]{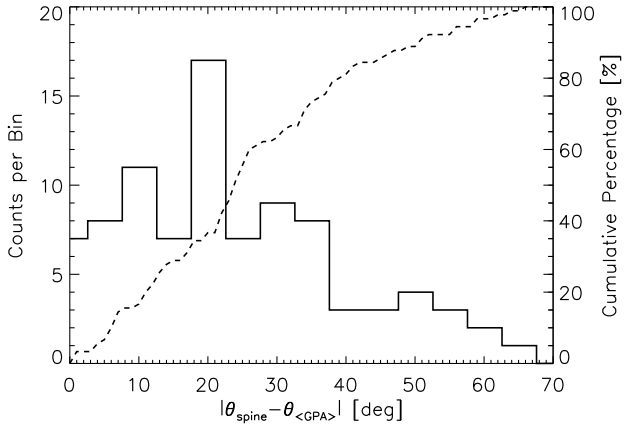}
\caption{Histogram of the absolute differences between the local spine orientation position angles and the local mean polarization Galactic position angles (GPA). The polarization position angles are primarily parallel to the spine (difference angles $<$ $35\,^{\circ}$) and not perpendicular, though some are oblique.}
\label{fig:paspine}
\end{figure}

Figure \ref{fig:paspine} examines the degree to which the direction of the magnetic field aligns with the cloud spine. It shows a histogram of the difference in the P.A. values between the local spine orientations and the local mean Galactic PA (GPA) polarization orientations. The mean polarization orientations were calculated by averaging the P.A.'s of vectors in non-overlapping 50 arcsec regions, centered on every other column spine location, to ensure independence. Therefore an individual stellar GPA could only be included in a single bin. The histogram shows the magnetic field is primarily parallel to the spine, with angle differences of 35$\degr$ or less being most common.

% Fig 6

\begin{figure}
\centering
\includegraphics[scale=0.39,angle=90]{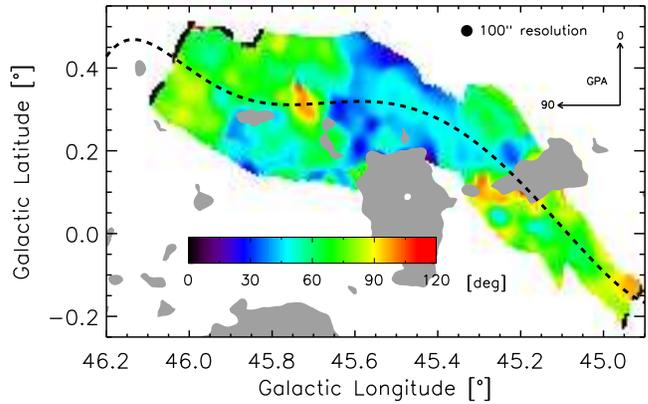}
\caption{Mean Galactic polarization position angle (GPA), spatially-averaged across the cloud. Gray regions represent areas where polarization vectors were removed, due to the potential for Cloud 2 contamination. The color bar represents GPA values from 0 to 120 degrees. The map has 100 arcsec resolution, as represented by the filled circle in the upper right.}
\label{fig:pamap}
\end{figure}

Additional maps were created to investigate the large-scale polarization properties of Cloud 1. GPIPS equatorial position angles for the polarization P.A.s were rotated to GPAs and averaged spatially, weighted by their uncertainties and by a gaussian offset weighting (see Sec. 4.2), to produce Figure \ref{fig:pamap}. Fig. \ref{fig:pamap} shows three distinct longitude-based regions. The regions on both ends of the cloud have GPA values similar to the spine at those locations. The middle region, however, shows GPAs at a significantly different value and more deviations from the spine angle.

% Fig 7

\begin{figure}[b]
\centering
\includegraphics[scale=0.39,angle=90]{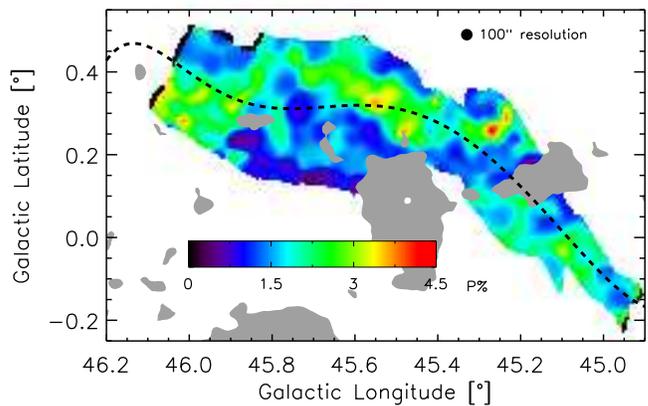}
\caption{Map of the spatially-averaged polarization percentage. The color bar represents polarization values from 0 to 4.5\%. The map has 100 arcsec resolution, as represented by the filled circle in the upper right.}
\label{fig:pmap}
\end{figure}

Figure \ref{fig:pmap} shows the mean degree of starlight polarization across the cloud, weighted in the same way as the GPA map. High polarization values are seen in regions corresponding to high $^{13}$CO integrated intensities, but Fig. \ref{fig:pmap} has few other large scale features. Both figures contain gray regions marking Cloud 2 exclusion zones.

% Fig 8

\section{Analysis}

With the physical and polarization characteristics for the cloud determined and analyzed, they were combined to estimate magnetic the plane-of-sky field strength across Cloud~1.

\subsection{Chandrasekhar-Fermi Method Application}

The CF method, as modified by \citet{OSG00} to return plane-of-sky field strengths, was used to estimate the magnetic field strength across Cloud 1. This method uses the observed polarization P.A. dispersions, the local gas density, and the local gas velocity dispersion to estimate the magnetic field strength, as:
\begin{equation}
B = 0.5 \: \Big{(} \frac{4}{3} \: \pi \: \rho \Big{)}^{0.5} \: \frac{\sigma_{v}}{\alpha}  \: \: \: \: \: [\mu G],
\end{equation}
where $B$ is the magnetic field strength in the plane-of-the-sky, $\rho$ is the volume mass density (in g cm$^{-3}$; Sec. 2.3), $\sigma_{v}$ is the $^{13}$CO gas velocity dispersion (in cm s$^{-1}$; Sec. 2.2), and $\alpha$ is the angular dispersion of the polarization vectors (in radians; Sec. 4.2). The volume mass density was calculated using Eq. (\ref{eq:five}) multiplied by molecular Hydrogen's weight and a factor of 1.36 to account for heavier elements \citep{SIM01}. The 0.5 coefficient was added by \citet{OSG00} for estimating plane-of-sky magnetic field strengths.

\subsection{Polarization Binning, Resolution, and Dispersion}

To estimate magnetic field strengths with the CF method, the P.A. dispersions ($\alpha$) across the cloud are also needed. Using the same 22 arcsec ($\ell$, $b$) grid characterizing GRS, for each grid pixel, all polarizations within a designated radial offset of each grid center were included in the P.A. dispersion calculations. These were computed using weighting by the P.A. uncertainties. This approach, however, created map artifacts, including hard boundaries at edges of the chosen search radii.

To reduce these effects, an additional distance-based gaussian weighting was adopted that reduced the weights of the P.A.s for stars located far from each grid center. This created smoother images, though at coarser angular resolution.

\begin{figure}
\centering
\includegraphics[scale=0.35,angle=90]{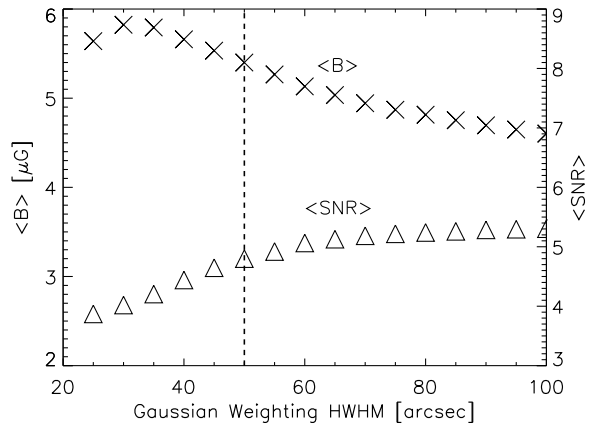}
\caption{Average Cloud 1 magnetic field strength (crosses) versus HWHM width of the gaussian weighting function applied to polarization position angles. Also plotted (open triangles) are the Cloud 1 average magnetic field SNRs. An increase in the SNR is seen up to about 60 arcsec. To balance resolution versus SNR, a compromise HWHM of 50 arcsec was chosen and is shown as the dashed vertical line.}
\label{fig:hwhm}
\end{figure}

\begin{figure}
\centering
\includegraphics[scale=0.39,angle=90]{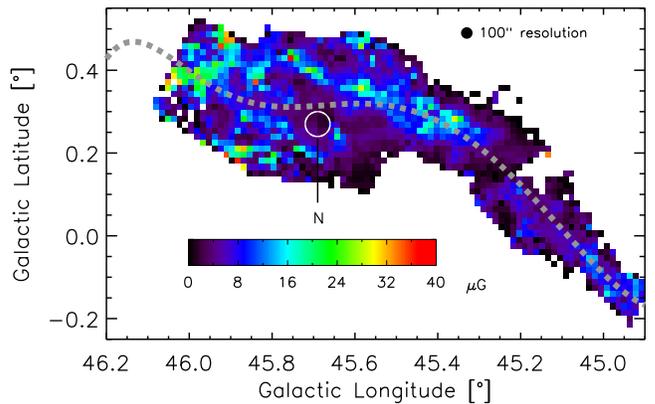}\caption{Magnetic field strength (plane-of-sky) for Cloud 1 (GRSMC 45.60+0.30), shown on a 50 arcsec grid and having 100 arcsec angular resolution. The color bar represents magnetic field strength values from 0 to 40 $\mu$G. Gray dashed curve traces the cloud spine, described in Sec 2.3. The white circle with `N' label is described in Sec. 5.1.}
	\label{fig:Bmag_paper}
\end{figure}

To choose the best gaussian width for weighting and averaging, magnetic field strength maps were produced using a range of these widths. Uncertainties were also propagated to create corresponding magnetic field strength uncertainty maps. The gaussian widths needed to allow enough stars to be included at each grid point to yield low uncertainty P.A. dispersions, but small enough to maintain good angular resolution. Figure \ref{fig:hwhm} shows the mean signal-to-noise ratio for the magnetic field strength ($<$SNR$>$, triangles) averaged across the cloud as a function of the gaussian weighting HWHM width. The signal-to-noise becomes flat beyond 60 arcsec. Fig. \ref{fig:hwhm} also shows the cloud average (plane-of-sky) magnetic field strength (crosses) as a function of weighting HWHM and illustrates the effect of oversmoothing; as the HWHM width grows, the average estimated mean magnetic field strength decreases. An HWHM width of 50 arcsec was chosen as a compromise value. This resulted in P.A. dispersions calculations using an average of $18\pm4$ stellar polarizations per $22 \times 22$\arcsec\, map pixel. To create a map of {\it independent} measurements after this nominal weighted averaging, the map was post facto resampled to a $50\times50$ arcsec (Nyquist) grid, to become Figure \ref{fig:Bmag_paper}.

% Fig 9

% Fig 10

\section{Results}

Using the NIR stellar polarizations, suitably binned and smoothed, with the $^{13}$CO GRS spectral line data, the CF method was applied to Cloud~1 to yield a full, resolved, plane-of-sky magnetic field strength map. In the following sections, this magnetic field strength map is characterized and its properties explored.

\subsection{Large-Scale Magnetic Field Characteristics}
Figure \ref{fig:Bmag_paper} presents the first resolved plane-of-sky magnetic field strength map across a complete molecular cloud. Throughout the cloud interior, the filamentary nature of Cloud~1, as seen in the gas, is mirrored by the ridge of enhanced magnetic field strength seen along the cloud spine. 

\begin{figure}
\centering
\includegraphics[scale=0.35,angle=90]{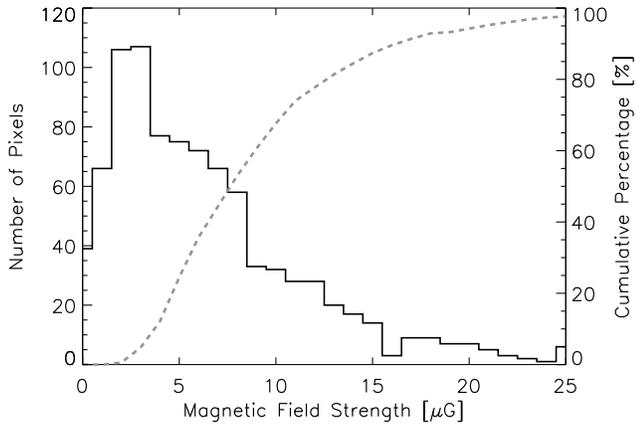}
\caption{Histogram of plane-of-sky magnetic field strengths (solid black) with SNR $\geq$ 3 for Cloud 1. Also plotted is the associated cumulative distribution (dashed gray curve). Median field strength is 7 $\mu$G.}
\label{fig:Histogram}
\end{figure}

Within Fig. \ref{fig:Bmag_paper}, 81\% of the magnetic field strength values have a SNR greater than 3 (902 individual measurements), and 92\% greater than 2 (1024 individual measurements). The weighted average magnetic field strength over the cloud is $5.40\pm0.04$~$\mu$G, the sample dispersion is 5.8 $\mu$G, and the maximum field strength is $44.3\pm13.5$ $\mu$G. The histogram and cumulative fraction of magnetic field strengths across the cloud are presented in Figure \ref{fig:Histogram}. The cumulative distribution function shows that 10$\%$ of magnetic field strengths are above 16~$\mu$G and 50$\%$ are above 7~$\mu$G. The mean magnetic field uncertainty is $1.8\pm0.5$~$\mu$G, with a dispersion of 1.4~$\mu$G. 

% Fig 11

\citet{OSG00} empirically determined that the CF method does not apply to regions with $\alpha$ $\geq25\degr$. One region (hereafter region `N'; see Fig. \ref{fig:Bmag_paper}), at ($\ell$, $b$)=(45.67,~0.29) and roughly 3 pc in diameter, violates this condition, and the estimated magnetic field strengths in this region are thereby upper limits. 

\subsection{$B$-$n_{H_2}$ Relation}

\begin{figure}
\centering
\includegraphics[scale=0.35,angle=90]{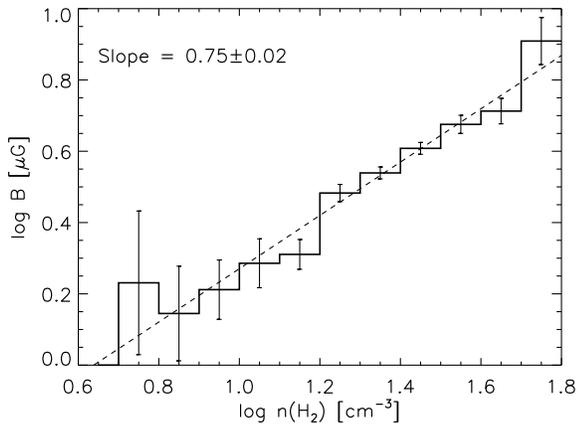}
\caption{Log-log plot of plane-of-sky magnetic field strength versus molecular hydrogen volume density, binned by 0.1 dex of gas density. The error bars represent $\pm$3$\sigma$ uncertainties. The dashed line is the least-squares fit $\log(B)\sim(0.75\pm0.02)\log(n_{H_2})$.}
\label{fig:bn}
\end{figure}

Comparing Figure \ref{fig:Bmag_paper} with Figure \ref{fig:numden}, a strong correlation was found between gas volume density (Fig. \ref{fig:numden}) and plane-of-sky magnetic field strength (Fig. \ref{fig:Bmag_paper}), as shown on the log-log plot in Figure \ref{fig:bn}. The magnetic strength data have been binned by gas density and averaged using weighting by their uncertainties. A linear least-squares fit returns a slope of $0.75\pm0.02$ for $log(B) / log(n_{H_2})$.

% Fig 12

\subsection{Magnetic Cores}

The magnetic field strength map in Fig. \ref{fig:Bmag_paper} reveals the presence of several isolated regions of high magnetic field strength, regions we designate `magnetic cores.' Such cores have not been seen before. In the following, we describe a method for their detection and delineation and study their properties.

\begin{figure}
\centering
\includegraphics[scale=0.39,angle=90]{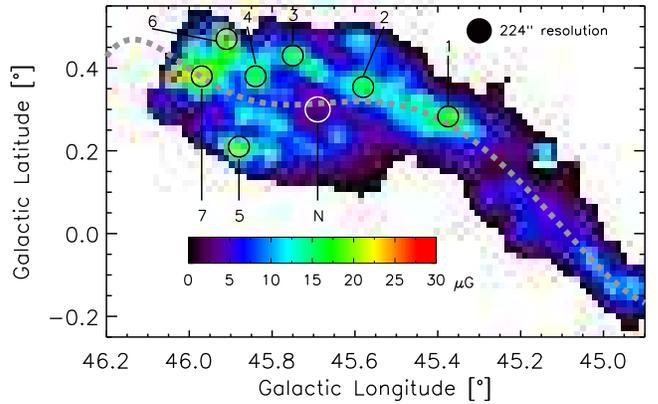}
\caption{Smoothed (plane-of-sky) magnetic field strength map, used for identifying magnetic cores. The seven magnetic cores and the `N' region are labeled. The color bar represents magnetic field strength values from 0 to 30 $\mu$G. The map has 224 arcsec resolution, as represented by the circle in the upper right.}
\label{fig:cores}
\end{figure}

To identify the cores, the magnetic strength image (Fig. \ref{fig:Bmag_paper}) was gaussian-smoothed using a HWHM of two pixels (100 arcsec) to create Figure \ref{fig:cores}. Multiple magnetic strength contours were overlaid, beginning at zero and stepped by 1.8 $\mu$G, the average magnetic field strength uncertainty ($\sigma_{B}$) of the unsmoothed map. The number of separate cores seen at each contour level was counted. As the contour level increased, the number of isolated cores rose, peaking at six cores within the 6, 7, and 8 contour levels ($10.8, 12.6, 14.4~\mu$G) before decreasing in number at higher contour levels. Within these three contours, the distinct regions of strong magnetic field strength became isolated from their surroundings. Isolation helped distinguish the cores from the elevated magnetic field strength values exhibited along the spine of the cloud. One core was found in the unsmoothed image (Fig.~\ref{fig:Bmag_paper}) to consist of two distinct strength-enhanced regions. This was confirmed at the ninth contour, which broke this initial core into its distinct parts. The bright feature at ($\ell$, $b$) $\sim$ (46\degr.02, $+$0\degr.39) in Fig. \ref{fig:cores} was found to be an edge effect and so was rejected. Therefore, the final magnetic core count was judged to be seven. 

% Fig 13

These seven cores are marked on the smoothed (plane-of-sky) magnetic field
strength map, shown as Fig. \ref{fig:cores}. Gaussian fits were used to find the centers and sizes of the magnetic cores in the map. The remaining core properties were derived from the unsmoothed map and are listed in Table \ref{tab_core}. The radii listed in Table \ref{tab_core} are the harmonic means of the column and row gaussian half-max widths found from the gaussian fits to the smoothed map. In Table~\ref{tab_core}, the `$R_{COR}$' column reports the core radii corrected for the effects of the image
smoothing.

The average maximum (unsmoothed, plane-of-sky) magnetic field strength for the seven magnetic cores is $35\pm2$~$\mu$G. Core 3 shows the largest maximum magnetic field strength, at $42\pm10$ $\mu$G. The mean magnetic field strength across all cores is $8.3\pm0.9$~$\mu$G. For comparison, the remaining non-core regions of Cloud 1 have an average magnetic field strength of $5.1\pm0.1$~$\mu$G. 
The average corrected core radius is $1.2\pm0.2$~pc, with an average mass of $129\pm24$~$M_{\sun}$. The masses were calculated using Eq. (4) from \citet{SIM01} and summing over the fitted radii.
The cores appear somewhat evenly spaced along the spine of Cloud 1, with an average intercore distance of $600\pm100$~arcsec ($5.5\pm0.9$~pc at~1.88 kpc) between neighboring core centers. 

Figs. \ref{fig:coint} and \ref{fig:cores} show that the $^{13}$CO-traced dense cloud cores match the locations of the magnetic cores. The polarization P.A. dispersions are small in each core and the mean GPAs fall into two domains, one at approximately GPA~=~$47\arcdeg\pm2\arcdeg$ and the other at GPA~=~$71\arcdeg\pm2\arcdeg$ (see Table \ref{tab_core} and Fig. \ref{fig:pamap}). There do not appear to be significant differences in physical properties among the cores.

\subsection{Mass-to-Flux Ratios}

Are these new `magnetic cores,' magnetically supported or are they collapsing? To answer this, we consider the gravitational potential and the strength of the magnetic field, through the mass to flux ratio \citep{CRUT9}:
\begin{equation} \label{eq:mtf}
\frac{\mathcal{M}}{\Phi} = \frac{M/\Phi_{B}}{(M/\Phi_{B})_{crit}} = 1.0 \times 10^{-20} N(H_{2})/\left|B\right|,
\end{equation}
as normalized by the critical ratio. Values of mass-to-flux above unity correspond to supercritical (gravity dominated) conditions and values below unity correspond to subcritical (magnetic dominated) conditions. As discussed in \citet{CRUT9}, these values do not take into account any turbulent energy present in the cloud or cores.

The quantity $B$ in Eq.~(\ref{eq:mtf}) is the full, 3-D magnetic field strength. However, magnetic field strengths measured in this work are plane-of-sky values. If we assume the line-of-sight magnetic field strength is zero, then the plane-of-sky strengths would be the full field strengths. For any other choice, the plane-of-sky field strengths represent {\it lower limits} to the true field strength. Therefore, by substituting the plane-of-sky magnetic field strengths for $B$ in Eq.~(\ref{eq:mtf}), the mass-to-flux ratios calculated will represent upper limits to the true mass-to-flux ratios.

\begin{figure}
\centering
\includegraphics[scale=0.39,angle=90]{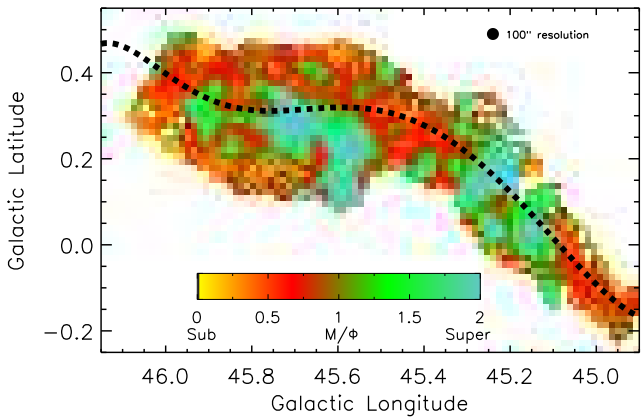}
\caption{Map of the (upper limits to the) mass-to-flux ratio across Cloud 1,
calculated by using the plane-of-sky field strengths in place of the full 
field strengths. Subcritical regions are seen along the cloud spine coincident with the magnetic cores. The color bar represents mass-to-flux upper limit values 
ranging from 0 to 2. The map has 100 arcsec resolution, as represented by the circle in the upper right.}
\label{fig:ftm}
\end{figure}

Mass-to-flux ratios were computed using the plane-of-sky magnetic field strengths substituted for $B$ for each pixel of the 50 arcsec resolution map, and are shown as Figure \ref{fig:ftm}. Eighty-two percent of the $M/\Phi$ values in the figure have SNR greater than three. The outer cloud envelope is too noisy to accurately characterize and makes up most of the low SNR population.

% Fig 14

In Fig. \ref{fig:ftm}, several extended, large-scale, supercritical regions are seen. These regions are larger than (but avoid) the magnetic cores and correspond to regions of low magnetic field strength but not necessarily low gas column density. 
The previously identified `N' region, where the C-F method may not apply, is one of the regions exhibiting a high mass-to-flux ratio, but this is really a lower limit since the magnetic field strength found there is an upper limit. 

All of the `magnetic cores' are subcritical (see last column in Table \ref{tab_core}), with a mean $M/\Phi$ of $0.74\pm0.04$, and so are magnetic dominated. They may be long-lived as a result of this magnetic support and the field would seem to be suppressing star formation in these cloud cores.

\section{Discussion}

This work presents the first resolved magnetic field map for an entire quiescent molecular cloud, in this case the 40 pc long cloud GRSMC 45.60+0.30. This map has an effective angular resolution of 100 arcsec (0.9 pc), a result made possible by the large number of GPIPS background starlight polarization measurements and coincident $^{13}$CO gas information. Combined with estimates of the cloud thickness and polarization position angle dispersion, the magnetic field was able to be estimated using the CF method for over 900 independent cloud directions. It is with these magnetic field estimates that we next explore relationships between magnetic field strength and gas density as well as between magnetic fields and cloud evolution.

\subsection{Magnetic Field Strength Dependence on Gas Density}

Within molecular clouds, the magnetic field strength $B$ is correlated with gas density $n_H$. \citet{TROL8} obtained a $B$-$n_H$ power law slope of 0.4 to 0.6 for their Zeeman work towards dark clouds at moderate gas densities ($> 100$~cm$^{-3}$). \citet{CRUT9} compiled a list of dark cloud magnetic field measurements and fit a power law index of 0.47$\pm$0.08 for densities greater than $1000$~cm$^{-3}$. Recently, \citet{CRU10} completed a comprehensive analysis of Zeeman measurements from the literature, attempting to quantify the actual magnetic field strength probability distribution function (PDF) by comparing the observed line-of-sight magnetic field strength PDF with different models. The {H\kern0.1em{\sc i}} Zeeman measurements are particularly relevant because they probe densities similar to the average density found here for Cloud 1. \citet{CRU10} concluded that the magnetic field is consistent with a two-part model wherein magnetic field strength is independent of n$_{H}$ below 300~cm$^{-3}$ and $log(B) / log(n_H) \approx 0.65$ 
above 300~cm$^{-3}$. 

\begin{figure}
\centering
\includegraphics[angle=90,scale=0.45]{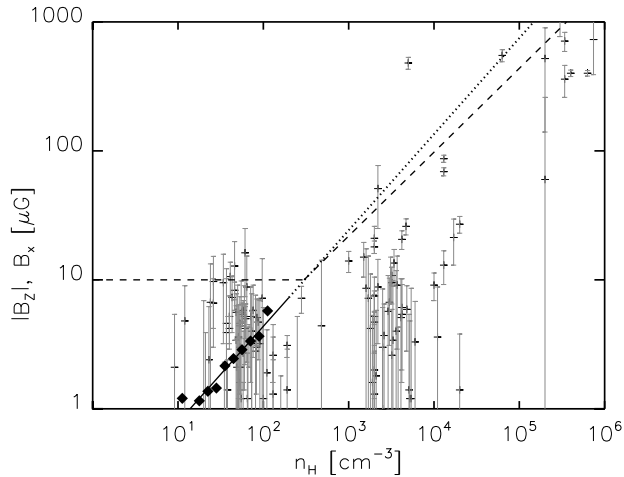}
\caption{Log-log plot of one-dimensional magnetic field strength against atomic hydrogen (equivalent) gas volume number density. Line-of-sight $B_Z$ data from \citet{CRU10} are shown as gray crosses and error bars. The \citet{CRU10} model is shown as the broken dashed line. Shown as filled diamonds are the plane-of-sky data from Fig.~\ref{fig:bn}, without error bars, shifted down by 0.15 dex to synthesize
one-dimensional $B_X$ values and shifted right by 0.3 dex to count $H_2$
densities twice, as per \citet{CRU10}. The solid black line is the similarly shifted fit to the plane-of-sky data, with dotted extension beyond the fit region. Note how the new data  depart from the model
below $n_H \sim 300$ cm$^{-3}$ but predict good agreement for higher
densities.}
\label{fig:cntb}
\end{figure}

The relationship between plane-of-sky magnetic field strength and molecular hydrogen density for Cloud~1 was presented as Fig.~\ref{fig:bn}. It showed a $B$-$n_{H_2}$ relation power law index of $0.75\pm0.02$. The Fig.~\ref{fig:bn} results were combined with the \citet{CRU10} observational data (gray data and error bars) and their derived model (dashed black line) to become Figure~\ref{fig:cntb}. The \citet{CRU10} data are one-dimensional line-of-sight $B_Z$ values. The Cloud~1 data are the two-dimensional plane-of-sky measurements reduced along the y-axis by root two to synthesize one-dimensional $B_X$ (or $B_Y$) values and shifted to positive $n_H$ by 0.3 dex to convert Cloud 1 H$_2$ densities to {H\kern0.1em{\sc i}}, as was done
by \citet{CRU10}. 

% Fig 15

The trend seen for Cloud 1 is inconsistent with the constant magnetic field
strength below 300 cm$^{-3}$ in the model of \citet{CRU10}. Previous work had suggested that below some threshold gas density (between 100 and 300 cm$^{-3}$), the $B$-$n_H$ relationship broke down \citep{TROL8,CRU10}. The Cloud 1 results presented here show that the $B$-$n_H$ relationship survives to the low densities probed in GRSMC 45.60+0.30, and with an index nearly identical to the one found at higher densities by \citet{CRU10}.

\subsection{Cloud and Core Stability and Evolution}

The role of magnetic fields in the support of GRSMC 45.60+0.30 was revealed by the predominance of subcritical values for the mass-to-flux ratios, calculated in Sec. 5.4. In Fig. \ref{fig:ftm}, many areas of Cloud~1 are magnetically-supported against collapse, including the high-density cores. Importantly, since the mass-to-flux values are upper limits, all subcritical designations are robust.

All of the magnetic cores are subcritical; the magnetic fields likely provide important magnetostatic support. Though the effects of turbulence were not considered, including them would only {\it reduce} the magnetic support necessary to prevent collapse, making the cores even more stable against gravity.

The magnetic cores are coincident with the high column density regions seen in the $^{13}$CO integrated intensity map (Fig. \ref{fig:coint}). The arrangement of the magnetic cores along the cloud's spine of more general $^{13}$CO emission and the overall alignment of the spine with the projected magnetic field (Fig. \ref{fig:paspine}) suggest that the magnetic field may be important in the formation and evolution of both the cloud and the cores. The enhancement of the magnetic field in the high-density regions suggests that the magnetic field has been amplified as the cores formed from more diffuse intercore gas. The average magnetic field strength for the cores is about 1.5 times the mean magnetic field strength for the cloud ($8.3\pm0.9$ $\mu$G versus $5.40\pm0.04$ $\mu$G). The core mean plane-of-sky magnetic field strength is also higher than the value typically found in the cold HI phase of the ISM \citep[6~$\mu$G for the full 3-D field strength;][]{HEIL5}. No correlation was found between gas number density in the cores and their level of mass-to-flux criticality.

These cores would therefore seem to be the best example of magnetic fields, even weak ones, regulating the star formation process. It is noteworthy that 8~$\mu$G magnetic fields can prevent these $\sim130$~M$_{\sun}$ dense cores from collapsing for what must be at least several free-fall times.

\subsection{Characteristic Core Spacing}

The existence of magnetic cores that are coincident with regions of high $^{13}$CO density and that are relatively uniformly spaced may relate to theories of fragmentation for a self-gravitating fluid cylinder \citep[``sausage'' instability]{CF53b,N87}. Following the approach of \citet{JAC10}, Cloud~1 could be considered to be either an incompressible fluid or an infinite isothermal gas cylinder, yielding two distinct characteristic core spacings. For the conditions present in Cloud~1, these spacings are 13 and 17 pc, respectively. They are both larger than the measured mean core spacing of $5.7\pm0.9$~pc. Because of the low volume densities present in Cloud~1, the gravitationally-driven ``sausage" instability is unlikely to account for the observed core spacing. The Parker instability \citep{PAR66} was also considered, but the expected instability scale length is also too large to account for the Cloud's core spacing.

\section{Summary}

By combining new NIR starlight polarimetry from GPIPS with $^{13}$CO data from GRS, and photometry from 2MASS, the CF method was used to measure the plane-of-sky magnetic field strength across the quiescent molecular cloud GRSMC 45.60+0.30 (Cloud 1). This study yielded the following:
\begin{itemize}
	\item The plane-of-sky magnetic field within Cloud 1 is stronger and more ordered in the densest regions of the cloud and weaker in the lower density regions. The average plane-of-sky magnetic field strength is $5.40\pm0.04$ $\mu$G, with a dispersion of 5.8 $\mu$G.
	\item Seven `magnetic cores' were identified within Cloud 1. Their average plane-of-sky magnetic field strength is $8.3\pm0.9$ $\mu$G, their average peak strength is $35\pm2$ $\mu$G, their average radius is $1.2\pm0.2$ pc, and their average mass is $129\pm24$~M$_{\sun}$. 
	\item The average intercore distance is $5.7\pm0.9$ pc, with core-to-core distances spanning 3.5 pc to 8.2 pc. Neither the ``sausage" nor Parker instabilities can account for the short spacings seen in this cloud.
	\item The magnetic cores all have subcritical mass-to-flux ratios. The cores appear magnetically-supported against collapse, likely inhibiting star formation. This may be the best evidence to date for the vital role played by the magnetic field in regulating star formation.
	\item The power law index linking magnetic field strength and gas number density is $0.75\pm0.02.$ This work extends this relation to lower densities than probed directly by \citet{CRU10}, and this index is in disagreement with their predictions for $n_H < 300$ cm$^{-3}$, suggesting that the magnetic field strength and gas number density are related even in the least dense regions of clouds.

\end{itemize}

\subsection{Acknowledgments}
Meredith Bartlett, Julie Moreau, Katie Jameson, and especially April Pinnick provided input, discussion and suggestions. We thank Richard Crutcher for permission to reproduce his data. This publication makes use of molecular line data from the Boston University-FCRAO Galactic Ring Survey, a joint project of Boston University and Five College Radio Astronomy Observatory, funded by the NSF under grants AST-9800334, AST-0098562, AST-0100793, AST-0228993, \& AST-0507657. This research was conducted in part using the Mimir instrument, jointly developed at Boston University and Lowell Observatory and supported by NASA, NSF, and the W.M. Keck Foundation. This work is partially supported by NSF grants AST-0440936 and AST-0607500 (PI Dan Clemens), Boston University's continuing Perkins Telescope partnership with Lowell Observatory, Boston University's Undergraduate Research Opportunities Program and Federal Work-Study. 

{\it Facilities:} \facility{Perkins}

\newpage

%\begin{landscape}
\begin{deluxetable}{cccccccccccccc}
\tabletypesize{\scriptsize}
%\rotate
\tiny
\setlength{\tabcolsep}{0.06in}
\tablecolumns{14} 
\tablewidth{0pt} 
\tablecaption{Magnetic Core Properties}
\tablehead{\colhead{Core}&\colhead{$\ell$}&\colhead{b} &\colhead{N$_{P}$\,\tablenotemark{a}} &
   \colhead{$<\overline{GPA}>$}&\colhead{$\alpha$}&
   \colhead{$<\sigma$$_{V}>$}&\colhead{$<n_{H_{2}}>$}&\colhead{Mass}&\colhead{B$_{max}$}&
   \colhead{$<B>$}&\colhead{Radius}&\colhead{R$_{COR}$\,\tablenotemark{b}}&\colhead{$\overline{M/\Phi}$}\\
 \colhead{}&\colhead{[$\deg$]}&\colhead{[$\deg$]}&\colhead{}&\colhead{[$\deg$]}&\colhead{[\phn{\arcdeg}]}& 
   \colhead{[km s$^{-1}$]}&\colhead{[cm$^{-3}$]}&\colhead{[$M_{\sun}$]}&\colhead{[$\mu$G]}&               
   \colhead{[$\mu$G]}&\colhead{[pc]}&\colhead{[pc]}&\colhead{}}   
\startdata
 1  & 45.38 & 0.24 & 169 & 48.6$\pm$0.2 & 11.2$\pm$0.2 & 
 0.83$\pm$0.03 & 44.8$\pm$1.5 & 162$\pm$3  & 30$\pm$4  & 7.7$\pm$0.2  & 1.77$\pm$0.05 & 1.45 & 0.78$\pm$0.04 \\
 2  & 45.60 & 0.32 & 68 & 45.9$\pm$0.4 & 10.8$\pm$0.4  &  
1.08$\pm$0.03 & 33.5$\pm$1.7 & 121$\pm$2  & 30$\pm$7  & 10.7$\pm$0.4 & 1.14$\pm$0.02 & 0.51 & 0.92$\pm$0.06 \\
 3  & 45.75 & 0.39 & 91  & 66.1$\pm$0.3 & 12.9$\pm$0.3 &  
1.36$\pm$0.10 & 27.6$\pm$1.2 & 118$\pm$3 & 42$\pm$10  & 6.7$\pm$0.2 & 1.45$\pm$0.09 & 1.03 & 0.79$\pm$0.07 \\
 4  & 45.85 & 0.35 & 63  & 71.9$\pm$0.5 & 13.7$\pm$0.5 & 
1.43$\pm$0.09 & 26.2$\pm$1.7 & 55$\pm$2  & 37$\pm$6  & 10.1$\pm$0.6  & 1.02$\pm$0.02 & 0.50\,\tablenotemark{c} & 0.63$\pm$0.07 \\
 5  & 45.89 & 0.17 & 87  & 49.5$\pm$0.3 & 23.4$\pm$0.3 &  
1.15$\pm$0.07 & 40.5$\pm$10.2 & 78$\pm$2  & 38$\pm$12  & 5.0$\pm$0.3  & 1.72$\pm$0.08 & 1.38 & 0.64$\pm$0.07 \\
 6  & 45.92 & 0.42 & 81  & 69.2$\pm$0.3 & 15.4$\pm$0.3 & 
1.61$\pm$0.10 & 29.0$\pm$1.5 & 154$\pm$3  & 35$\pm$9 & 7.0$\pm$0.3  & 1.95$\pm$0.10 & 1.66 & 0.67$\pm$0.04 \\
 7  & 45.99 & 0.35 & 95  & 74.7$\pm$0.2 & 19.1$\pm$0.2 & 
1.88$\pm$0.10 & 38.9$\pm$2.0 & 218$\pm$5  & 32$\pm$6  & 10.9$\pm$0.3  & 1.85$\pm$0.05 & 1.54 & 0.74$\pm$0.06 \\		   
		   \hline
\multicolumn{3}{l}{Unweighted Means}& 93& 61& 15& 
1.33& 34& 129& 35& 8.3& 1.6& 1.2& 0.74\\
\multicolumn{3}{l}{Uncertainties \,\tablenotemark{d}}& 14&  5&   2& 
0.14&   3&   24&   2& 0.9& 0.2& 0.2& 0.04\\
\enddata
\label{tab_core}
\tablenotetext{a}{Number of polarization measurements within each core.}
\tablenotetext{b}{Corrected for the 1.02 pc (224\arcsec) smoothing gaussian.}
\tablenotetext{c}{Unresolved, set to half the smoothing gaussian size.}
\tablenotetext{d}{Computed from standard deviations, reduced by $\sqrt{6}$.}
\end{deluxetable}
%\clearpage
%\end{landscape}

\end{document}